\documentclass[12pt]{article}
\usepackage[latin1]{inputenc}
\usepackage{graphicx}
\usepackage{amssymb}

\newcommand{\Ncal}{{\cal N}}
\newcommand{\drm}{{\rm d}}

\newcommand{\pa}{\partial}

\newcommand{\text}{\rm}

\newcommand{\ug}{ \; = \; }

\newcommand{\bb}{\begin{equation}}
\newcommand{\ee}{\end{equation}}
\newcommand{\bega}{\begin{eqnarray}}
\newcommand{\ega}{\end{eqnarray}}
\newcommand{\begae}{\begin{eqnarray*}}
\newcommand{\egae}{\end{eqnarray*}}

\newcommand{\h}{\hspace*{4ex}}
\newcommand{\dis}{\displaystyle}

\newcommand{\om}{\omega}

\newcommand{\cent}{\centerline}
\newcommand{\vs}{\vspace*}

\begin{document}

\baselineskip 0.8cm

\begin{center}

{\Large {\bf Carving light beams${^\dag }$ }}
\footnotetext{$^{\: ^\dag}$  Work supported by FAPESP and CNPq. \ E-mail addresses for
contacts: mzamboni@decom.fee.unicamp.br}

\end{center}

\vs{2mm}

\cent{Michel Zamboni-Rached}

\vspace*{0.2 cm}

\cent{{\em School of Electrical and Computer Engineering, University of Campinas, Campinas, São Paulo, Brazil}}


\vs{0.5 cm}

{\bf Abstract  \ --} \ Some years after the appearance of the so-called non-diffracting beams, there was the development of methods capable of structuring them spatially, being the so called Frozen Waves method the first and, perhaps, the most efficient one. That method allowed modelling the longitudinal intensity pattern of non-diffracting beams, being, however, little efficient in controlling their transverse spatial pattern, granting only the possibility of choosing their transverse dimensions, which remain invariant throughout the propagation. In this work, we have extended the Frozen Wave method in such a way to control, in addition to the longitudinal pattern, the transverse beam structure along the propagation.
The new transversally and longitudinally structured beams can have potential applications in areas such as photonics, optical manipulation, optical atom guidance, lithography, etc..


\section{Introduction}

\h The theory of Localized Waves, also called Non-diffracting Waves (beams and pulses), had a diffuse beginning with a series of pioneering works by several researchers and research groups \cite{shep,brit,durnin,sez,besi1,lu,donn,recami1,saari,livro1,livro2}, resulting in new families of solutions for the wave equation and also for Maxwell's equations, representing beams and pulses immune to the diffraction effects or, in the case of finite energy solutions, resistant to its effects over long distances when compared to the ordinary waves. More specifically, a non-diffracting beam or pulse keeps its transverse spatial structure and intensity along the propagation.

\h It was thinking in bringing a greater flexibility to the structure of the new beams that the so-called Frozen Wave (FW)
method was developed \cite{fw1}, allowing the control of the longitudinal intensity pattern of non-diffracting Bessel-type beams. This approach is based on the superposition of co-propagating Bessel beams of the same order and frequency, and whose complex amplitudes and longitudinal wave numbers are chosen appropriately in order to obtain the desired longitudinal intensity modelling which, in the case of FWs generated by zero-order Bessel beams (zero-order FW beams), will be concentrated on the propagation axis and with a previously chosen spot radius, or, if generated by higher order Bessel beams (higher order FW beams), will be concentrated on a cylindrical surface (i.e., a hollow beam) of previously chosen radius \cite{fw1,fw2,fw3}.

\h In spite of providing a great control over the beam's intensity along the propagation (let us call such control of
"longitudinal control"), the method allows a much more limited control over the beam's transverse structure, allowing us to set its spot radius or its cylinder radius, and these remain invariant throughout the propagation.

\h In this paper, by exploring a degree of freedom present in the FW method, we extended it in such a way to get, in addition
to a strong longitudinal spatial control, a transverse one much more effective. With this, we will able to control the zero-order FW beam spot radius or the radius of the higher-order (hollow) FW beam along the propagation, bringing a very interesting and promising malleability to the new beams, which become candidates for potential applications in areas like photonics, optical manipulation, optical atom guidance, lithography, etc..

\h In Sec. II, we will introduce the extended method and in Sec. III we will apply it to some examples.
Section IV is dedicated to the conclusions.

\section{The method}

\h In the FW method \cite{fw1,fw2,fw3}, we require that the optical beam, $\Psi(\rho,\phi,z,t)$, to be constructed possesses, within $0 \leq z \leq L$, a predefined longitudinal intensity pattern either over the propagation axis ($z$ axis), with a given spot radius $r_0$, in the case of a needle beam, or over a cylindrical surface of radius $\rho_{\nu}$, in the case of a hollow beam; both radii also chosen in advance. In mathematical terms, we demand that $|\Psi(\rho=\rho_{\nu},\phi,z,t)|^2 \approx |F(z)|^2$ within $0 \leq z \leq L$, where $F(z)$ is called morphological function and its modulus square yields the desired longitudinal intensity pattern. For a needle beam, we write, by definition, $\rho_0 = 0$ and we choose the value of its spot radius $r_0$, while that for a hollow beam ($\nu \neq 0$) we choose the value of its transverse radius $\rho_{\nu}$.

\h According to the FW method, for obtaining a beam with such characteristics, we write it as a superposition of $2N+1$ Bessel
beams of order $\nu$:

    \begin{equation} \label{FW1}
        \Psi(\rho,\phi,z,t) \ug \Ncal_{\nu}\,e^{-i\omega t}\sum^{N}_{n=-N}A_{n}
        J_{\nu}(h_n\rho)e^{i\nu\phi}e^{i\beta_n z}\,\,,
        \end{equation}
        with the longitudinal wavenumbers given by

        \bb \beta_n \ug Q + \frac{2\pi}{L}n \,\,,  \label{betan}\ee
        and the transverse wavenumbers, as demanded by the wave equation,

        \bb h_n \ug \sqrt{k^2 - \beta_n^2} \,\,. \label{hn}\ee

        Also according to the method, the complex amplitudes are chosen as

        \begin{equation} \label{An}
        A_{n} = \frac{1}{L} \int^{L}_{0}F(z) e^{-i\frac{2\pi}{L}n\,z}
        \drm z\,\, .
        \end{equation}

\h In Eq.(\ref{FW1}), $\Ncal_{\nu} = 1/[J_{\nu}(.)]_{max}$, where
$[J_{\nu}(.)]_{max}$ is the maximum value of the Bessel function
of the first kind $J_{\nu}(.)$.

\h The value of $Q$ in Eq.(\ref{betan}) is chosen according to the desired spot radius, $r_0$, to the zero-order ($\nu=0$)
FW beam through

    \bb r_0 \approx \frac{2.4}{\sqrt{k^2 - Q^2}}\,\,,
    \label{r0}\ee
    or according to the desired radius, $\rho_{\nu}$, of the
    hollow-FW-beam through

    \bb \rho_{\nu} \approx \frac{\eta_{\nu}}{\sqrt{k^2 - Q^2}}\,\,,
    \label{rhonu}\ee
    where $\eta_{\nu}$ is such that $J_{\nu}(\eta)$ has its maximum value at $\eta=\eta_{\nu}$.

\h As it has been said, despite offering a high degree of control over the beam longitudinal behavior,  the FW method
is much more limited with respect to the control of the beam transverse characteristics. Now, we are going to overcome such limitation.

\h Due to the fact that the FW method is exclusively focused on the beam intensity envelope, the morphological function
$F(z)$, which provides the desired longitudinal intensity $ |F(z)|^2$, is, in general, a real function with very little or no phase variation. Now, it is clear that if instead of choosing a function $ F(z) = f(z) $, with $f(z)$ real, we choose

\bb F(z) \ug f(z)\exp(i g(z))\,\,, \label{F} \ee
with $f(z)$ and $g(z)$ real functions, we will still have the same desired longitudinal intensity pattern and, so, the phase function $g(z)$ can be understood as a degree of freedom within the FW method.

\h However, as we will see in the following, this very same phase function plays a fundamental role in the evolution of the
transverse pattern of the beam along the propagation.

\h When the morphological function $F(z)$ has a phase with no (or very little) variation, the parameter $Q$ is, in general,
the central value of the longitudinal wave numbers, Eq.(\ref{betan}), of the Bessel beams that constitute the FW-beam, Eq. (\ref{FW1}), also determining its spot radius $r_0$, in the case of a zero-order FW-beam, or its transverse radius $\rho_{\nu}$, in the case of the higher-order FW-hollow-beam, through Eqs.(\ref{r0},\ref{rhonu}).

\h Let us now consider the case where the morphological function is of the form given by Eq.(\ref{F}). By using
Eqs.(\ref{FW1},\ref{betan},\ref{An}) and considering, for simplicity, $\nu = 0$, we can write the solution for the FW-beam on the $z$ axis , i.e., for $\rho = 0$, as

    \begin{equation} \label{FW2}
        \Psi(\rho=0,z,t) \ug e^{-i\omega t}e^{i Q z}\sum^{N}_{n=-N}A_{n}
        e^{i\frac{2\pi}{L}n\,z} \approx e^{-i\omega t}e^{i Q z}f(z)e^{i g(z)}\,\,.
        \end{equation}

\h Within the neighborhood of a given $ z = z_0 $, i.e., for $ z = z_0 + \Delta z $, with \emph{small} values for $ \Delta z $, we can write

        \begin{equation} \label{FW3}
        \Psi(\rho=0,z_0+\Delta z,t) \, \approx \, \left[f(z_0)\,e^{i\left(Q z_0 + g(z_0)\right)}\right]\,e^{-i\omega t}\,e^{i\left(Q + \pa g/\pa z|_{z_0}   \right)\Delta z}\,\,,
        \end{equation}
        where we consider $ f(z_0 + \Delta z) \approx f (z_0) $ due to the small values assumed by $ \Delta z $.

\h From Eq.(\ref{FW3}), we see that, unlike a FW beam whose morphological function has no phase variation, this zero-order FW
beam will have, on the $z$ axis (where the field is concentrated), a fast variation given not just by $\exp(i Q\Delta z)$, but by $\exp({i(Q + \pa g / \pa z |_ {z_0}) \Delta z}$, which implies a beam spot radius given by $ r_0 \approx 2.4 / \sqrt{k^2 - (Q + \pa g / \pa z |_{z_0})^2} $ instead of that given by Eq.(\ref{r0}).

\h Since $z_0$ is a generic point, we can say that the spot radius of the zero-order FW beam will be a function of the
coordinate $z$:

    \bb r_0(z) \approx \dis{\frac{2.4}{\sqrt{k^2 - \left(Q + \dis{\frac{\pa g}{\pa z}}\right)^2}}}\,\,\,\,\,({\rm for}\,\,\nu=0)
    \label{r0z}\ee

\h In the case of higher-order FW beams, we can follow a similar reasoning, but with a little more care, and come to the
conclusion that the transverse radius of the hollow-FW-beam will also be a function of z:

\bb \rho_{\nu}(z) \approx \dis{\frac{\eta_{\nu}}{\sqrt{k^2 - \left(Q + \dis{\frac{\pa g}{\pa z}}\right)^2}}}\,\,\,\,\,({\rm for}\,\,|\nu|\geq 1)
    \label{rhonuz}\ee

\h Equations (\ref{r0z},\ref{rhonuz}) are fundamental, since they show that the degree of freedom in choosing the phase
function, $g(z)$, to the morphological function, Eq.(\ref{F}), can be used to control the spot radius of zero order FW beams or the transverse radius of the higher-order FW beams (which are hollow beams) along the propagation. Actually, from Eqs.(\ref{r0z},\ref{rhonuz}), we have that

\bb
 g(z) \ug \left\{\begin{array}{clr}
 -Q z + \dis{\int} \sqrt{k^2 - \dis{\frac{2.4^2}{r_0^2(z)}}}\, \drm z \;\;\; & {\rm for}\;\;\; \nu=0  \\
 \\
 -Q z + \dis{\int} \sqrt{k^2 - \dis{\frac{\eta_{\nu}^2}{\rho_{\nu}^2(z)}}}\, \drm z \;\;\; & {\rm for}\;\;\; |\nu|\geq 1
\end{array} \right. \,\,,\label{g}
 \ee
which can be used to calculate the necessary phase function $g(z)$ once the desired $r_0(z)$ or $\rho_{\nu}(z)$ have been chosen.

\h In this way, the now extended FW method can be announced as follows.

\h Let an optical beam be given by a superposition of $ 2N + 1 $ co-propagating Bessel beams of the same angular frequency, $\om$, and same order $\nu$, mathematically described by Eq.(\ref{FW1}). In such a case, it is possible to choose, on demand, the longitudinal intensity pattern the beam will acquire in a given range along the propagation as well as the evolution of its spot radius (if $\nu = 0$), or the evolution of the radius of the cylinder (if $|\nu| \geq 1$) on which the beam will be concentrated (hollow beam). Mathematically, in the range $0 \leq z \leq L$, in order to have $ |\Psi(\rho=\rho_{\nu}(z),\phi,z,t)|^2 \approx |f(z)|^2$, with $\rho_{\nu}(z)$ (for $|\nu| \geq 1$), $r_0 (z)$ (the spot radius for $\nu = 0$) and $f(z)$ \textbf{choosen at will}, the following choices have to be made: Eq.(\ref{betan}) for the longitudinal wave numbers, $\beta_n$, with the transverse ones, $h_n$, given by Eq.(\ref{hn}), the coefficients $A_n$ given by Eq.(\ref{An}), where $F(z)$ is now given by Eq.(\ref{F}) and the phase function $g(z)$ must be calculated through Eq.(\ref{g}), from the $\rho_{\nu}(z)$ or $r_0(z)$ desired.

\h Before proceeding to the examples, it is worth mentioning some important observations about the extended FW method:

\begin{itemize}
  \item{since the optical field is treated here as a scalar, the method must always be applied within paraxial situations, which implies $h_n << \beta_n$ for any $n$, or, similarly, we must have $ r_0 (z ) >> \lambda$ (for $\nu = 0$), $ \rho_{\nu}(z) >> \lambda$ (for $ |\nu| \geq 1 $) and $ L >> \lambda $.}
  \item{the FW method constructed from a discrete superposition of Bessel beams, as presented here, is more suitable for modeling beams within longitudinal regions of the order of centimeters or greater. For much smaller longitudinal spatial regions, micrometer regions, for instance, it is more appropriate to use the FW method based on continuous Bessel beam superposition \cite{fw4}. The extended method in such cases is obtained in a way similar to what we have done here, remembering that the vector character of the field must be taken into account.}
  \item{once the values of $L$ and $Q$ have been chosen, the value of $N$ in the summation of Eq.(\ref{FW1}) must be $N \leq (kQ) L/ 2\pi $ (because we must have $\beta_n <k$). Such restriction on the maximum value that $N$ can assume must also be observed as a limit on the accuracy with which the morphological function $ F (z) = f (z) \exp(ig(z)) $ can be represented by the truncated Fourier series $\sum_{-N}^{N} A_n \exp(i 2 \pi n z/L)$ and, therefore, the accuracy with which the resulting FW beam can represent the desired optical field.}
  \item{another important restriction can be obtained from Eq.(\ref{g}) and the imposition of working within the paraxial regime, from where we can infer that the phase function has to be such that $|\pa g/ \pa z| < (k-Q)$.  }

\end{itemize}

\section{Applying the method}

\h In this section we apply the extended FW method for obtaining some new and interesting beams whose longitudinal and transverse behavior along the propagation can be chosen at will.

\h In all cases, it is considered $\lambda = 0.532 \mu$m, widely used
in many applications.

\

\textbf{First Example:}

\h Here, we are interested in obtaining, within the spatial range $ 0 \leq z \leq L = 2 $cm, a hollow beam of uniform intensity that starts at $ z = z_i = 0.2 $ cm and whose radius (hollow beam radius) oscillates between the values $\rho_{\rm min}=4.3 \mu$m and $\rho_{\rm max}=13.3 \mu$m, with a spatial period of $\Lambda = 0.5$cm,  till it reaches $z=z_f=1.8$cm, from where the beam shall dissolve, acquiring negligible intensity.

\h We will proceed by considering our solution (\ref{FW1}) with $\nu=2$, so it is a second-order FW beam. Mathematically, for
obtaining a beam with the above characteristics, we can choose, within $ 0 \leq z \leq L = 2 $cm,

\bb
 f(z) \ug \left\{\begin{array}{clr}
 1 \;\;\; & {\rm for}\;\;\; 0.2{\rm cm} \leq z \leq 1.8 {\rm cm}
 \\
0 & {\rm otherwise} \,\,\,,
\end{array} \right. \label{f}
 \ee
and the desired hollow beam radius evolution along the propagation as

\bb \rho_{2}(z) \approx \dis{\frac{\eta_2}{\sqrt{k^2 - \left[Q + ab\cos\left(b\left(z-\frac{L}{2}\right)\right)\right]^2}}}\,\,\,,
    \label{rho2}\ee
    where we use that $\eta_{2} \approx 3.05 $, $Q=0.999 k$, $a=7.62$ and $b=1.26\times 10^3{\rm m}^{-1}$. From Eqs.(\ref{rhonuz},\ref{rho2}), we have that the phase, $g(z)$, of the morphological function, $F(z)=f(z)\exp(ig(z))$, has to be given by

    \bb g(z) \ug a\sin\left(b\left(z-\frac{L}{2}\right)\right) \,\,\, \label{g2} \ee

\h With $f(z)$ and $g(z)$ given by Eqs.(\ref{f},\ref{g2}), we have the complete morphological function, Eq.(\ref{F}), which is
used to calculate the coefficients $A_n$ of the main solution (\ref{FW1}) through Eq.(\ref{An}). Also, as it was seen in the previous section, the longitudinal and transverse wave numbers of the Bessel beam superposition (\ref{FW1}) are given by Eqs.(\ref{betan}) and (\ref{hn}), respectively. In this case we have $N=37$ and, so, $2N+1=75$ Bessel beams in that superposition.

 \begin{figure}[htbp]
\begin{center}
 \scalebox{.25}{\includegraphics{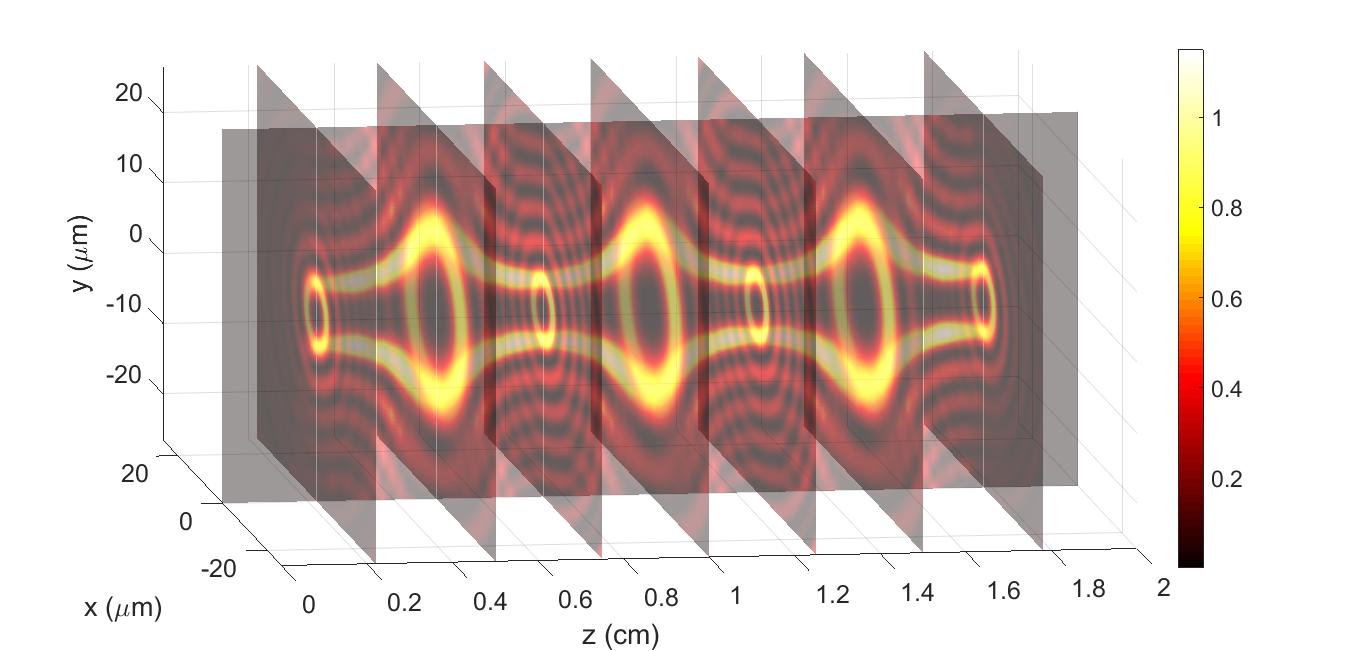}}
\end{center}
\caption{The 3D intensity of the resulting beam of the first example. We clearly see it possesses all the desired characteristics.} \label{fig1}
\end{figure}

\h Figure (\ref{fig1}) shows the 3D intensity of the resulting beam, which clearly possesses all the desired characteristics.

\h At this point, it is worth noting that, naturally, we could choose any other function $\rho_2(z)$ that yields the radius evolution described, in a general way, above Eq.(\ref{f}). It could be, for instance, $\rho_2(z) = (\rho_{\rm max}+\rho_{\rm min})/2 + (1/2)(\rho_{\rm max}-\rho_{\rm min})\cos(\Lambda(z-L/2))$, and $g(z)$ could be obtained through Eq.(\ref{g}). The only reason for choosing $\rho_2(z)$ as given by Eq.(\ref{rho2}) is simplicity, since with that it is mathematically much easier to obtain $g(z)$.

\

\textbf{More Examples:}

\h For sake of space, here we show four additional examples without going into their mathematical details and just describing the desired characteristics for each beam and showing the result obtained with the method. All the longitudinal distances involved in the examples are much greater than the diffraction distances of ordinary beams with similar transverse dimensions, which shows that the beams obtained here are resistant to diffraction effects, possessing transverse variations along the propagation that are chosen \emph{a priori}.

\h Figure (2a) shows a beam (not hollow one) designed to keep its intensity uniform within a given range, where its spot
radius was designed to have a behavior similar to the transverse radius of the hollow beam of the first example, i.e., to oscillate between a minimum and maximum value with a given spatial periodicity.

\h Figure (2b) shows a hollow beam that was designed to start at a given position, propagate a given distance while maintaining
its transverse radius, which then starts to oscillate along the propagation and, after some distance, return to its initial value, keeping it for an additional distance so, finally, the beam dissolves itself, acquiring negligible intensity values.

\h Figure (2c) shows a hollow beam designed to keep its intensity uniform within a given range, where its transverse
radius grows linearly with distance, giving the beam a light horn look.

\begin{figure}[htbp]
\begin{center}
\hspace*{-1.5cm}
\scalebox{.29}{\includegraphics{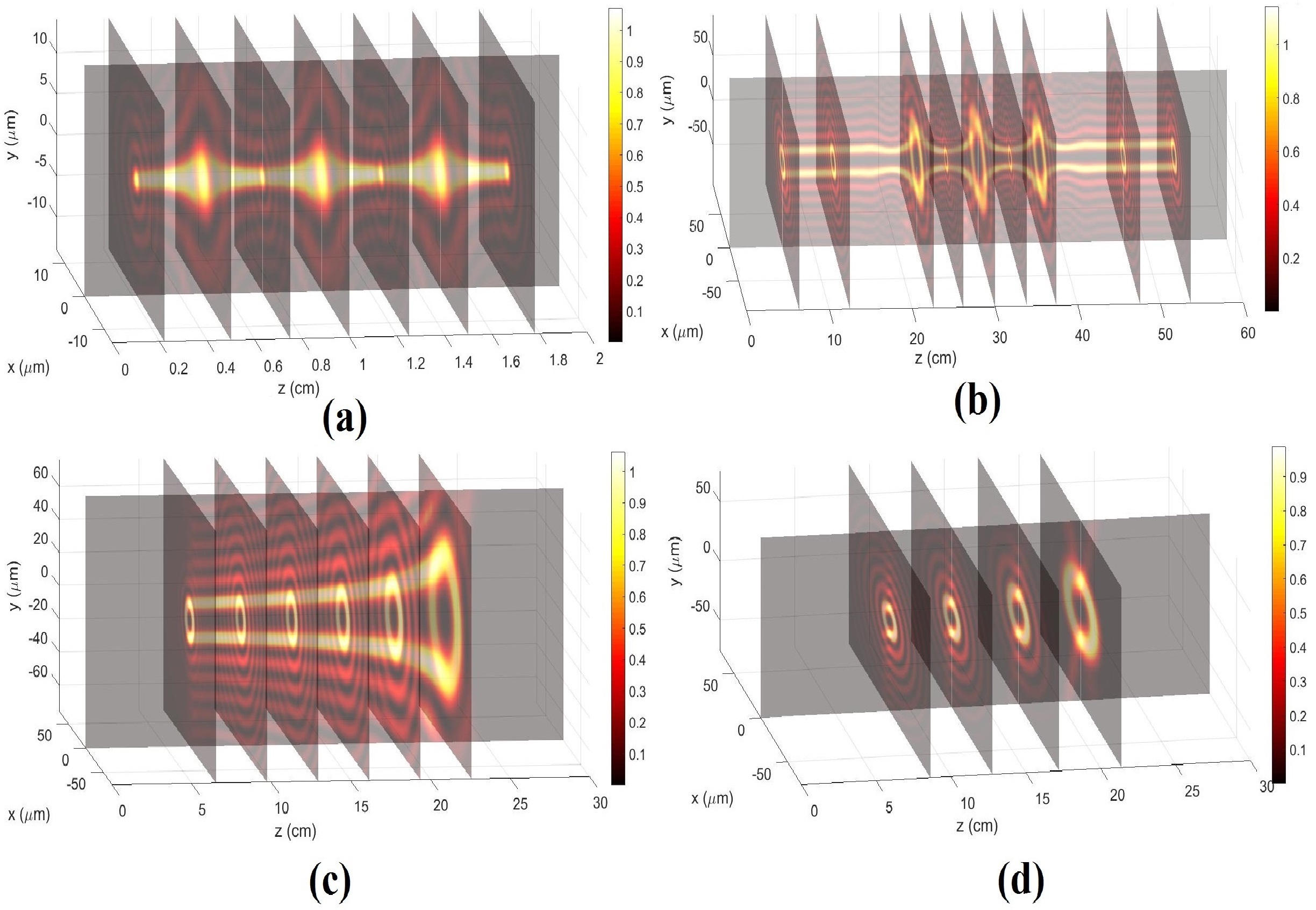}}
\end{center}
\caption{The 3D intensities of: (a) a not hollow beam designed to keep its intensity uniform within a given range, where its spot radius was designed to oscillate between a minimum and maximum values with a given periodicity; (b) a hollow beam designed to propagate a given distance keeping its transverse radius, which then starts to oscillate along the propagation to then return to its initial value; (c) a hollow beam designed to keep its intensity uniform within a given range and where its transverse radius grows linearly with distance; (d) a beam whose intensity was designed to assume an on-off behavior and whose transverse radius was designed to grow linearly with distance.} \label{fig2}
\end{figure}

\newpage

\h Finally, Fig.(2d) shows a hollow beam whose intensity was designed to assume an on-off behavior and whose transverse radius
was designed to grow linearly with distance, giving the beam a look of a sequence of light donuts of crescent radii.

\section{Conclusions}

\h In this work, by exploring a degree of freedom present in the Frozen Waves method, we extended it in such a way to control, in addition to the longitudinal pattern, the transverse beam structure along the propagation.
The new transversally and longitudinally structured beams can have potential applications in areas such as photonics, optical manipulation, optical atom guidance, lithography, etc..

\section*{Acknowledgements}

\h Thanks are due to partial support from FAPESP (under grant
2015/26444-8) and from CNPq (under grant 306689/2019-7). The
author also thanks Erasmo Recami for his continuous collaboration and interest.


\begin{thebibliography}{15}

\bibitem{shep} C.J.R. Sheppard and T. Wilson, ``Gaussian-beam theory of lenses
with annular aperture,'' Microwaves, Optics and Acoustics, Vol. 2,
No. 4 (1978).

\bibitem{brit} J.N.Brittingham, ``Focus wave modes in homogeneous Maxwell's equations: transverse
electric mode,'' J. Appl. Phys., Vol.54, pp.1179-1189 (1983).

\bibitem{durnin} J. Durnin, J. J. Miceli, and J. H. Eberly, ``Diffraction-free
beams,'' Phys. Rev. Lett., Vol. 58, pp. 1499-1501 (1987).

\bibitem{sez} A. Sezginer, ``A general formulation of focus wave modes,'' J.
Appl. Phys., Vol. 57, pp. 678-683 (1985).

\bibitem{besi1} I.M.Besieris, A.M.Shaarawi and R.W.Ziolkowski, ``A bi-directional
traveling plane wave representation of exact solutions of the
scalar wave equation," J. Math. Phys., Vol.30, pp.1254-1269
(1989).

\bibitem{lu} J.-y. Lu and J. F. Greenleaf, ``Experimental verification of
nondiffracting X-waves,'' IEEE Trans. Ultrason. Ferroelectr. Freq.
Control, Vol.39, 441-446 (1992).

\bibitem{donn} R.Donnelly and R.W.Ziolkowski, ``Designing Localized Waves,''
Proc. Roy. Soc. London, A, Vol.440, pp.541-565 (1993).

\bibitem{saari} Peeter Saari and Kaido Reivelt, ``Evidence of X-Shaped Propagation-Invariant Localized Light Waves,'' Phys.
Rev. Lett. 79, 4135 (1997).

\bibitem{recami1} E.Recami, ``On localized X-shaped superluminal solutions to Maxwell equations,''Physica A: Statistical
Mechanics and its Applications, Vol. 252, pp. 586-610 (1998).

\bibitem{livro1} For a more complete list of works and references, see also,
{\em Localized Waves}, edited by H.E.Hern\'andez-Figueroa, M.Zamboni-Rached, and E.Recami (J.Wiley; Hoboken,NJ, 2008).

\bibitem{livro2} For a more complete list of works and references, see also, {\em Non-Diffracting Waves}, edited by
H.E.Hern\'andez-Figueroa, E.Recami, and M.Zamboni-Rached
(J.Wiley; Berlin, 2014).

\bibitem{fw1} M.Zamboni-Rached, ``Stationary optical wave fields with arbitrary longitudinal shape by superposing equal
frequency Bessel beams: Frozen Waves,'' Opt. Express  $\mathbf{12}$(17), 4001--4006 (2004).

\bibitem{fw2} M.Zamboni-Rached, E.Recami, and H.E.Hern\'andez-Figueroa, ``Theory of `frozen waves'':
Modeling the shape of stationary wave fields,'' J. Opt. Soc. Am. A, Vo.22, pp. 2465-2475 (2005).

\bibitem{fw3} Michel Zamboni-Rached, "Diffraction-Attenuation resistant beams in
absorbing media," Opt. Express, Vol. 14, pp. 1804-1809 (2006).

\bibitem{fw4} Michel Zamboni-Rached, Leonardo André Ambrosio, Ahmed H. Dorrah, and Mo Mojahedi, "Structuring light under different polarization states within micrometer domains: exact analysis from the Maxwell equations," Opt. Express 25, 10051-10056 (2017)




\end{thebibliography}
\end{document}